\documentclass{Interspeech2024}

\usepackage{booktabs} 
\usepackage{array} 
\usepackage{multirow}
\usepackage{amsmath} 
\usepackage{arydshln}
\usepackage{graphicx}
\usepackage{hyperref}
\usepackage{cite}
\usepackage{amsmath}
\usepackage{comment}
\usepackage{xcolor}

\definecolor{deepblue}{rgb}{0, 0, 0.5}

\interspeechcameraready

\title{Understanding Sounds, Missing the Questions: The Challenge of Object Hallucination in Large Audio-Language Models}

\name[]{Chun-Yi}{Kuan}
\name[]{Wei-Ping}{Huang}
\name[]{Hung-yi}{Lee}

\address{
  Graduate Institute of Communication Engineering, National Taiwan University, Taiwan
}
\email{chunyi.kaun.tw@gmail.com, r11942102@ntu.edu.tw, hungyilee@ntu.edu.tw}

\keywords{Large audio-language models, Object hallucination}

\begin{document}

\maketitle

\begin{abstract}

Large audio-language models (LALMs) enhance traditional large language models by integrating audio perception capabilities, allowing them to tackle audio-related tasks. 
Previous research has primarily focused on assessing the performance of LALMs across various tasks, yet overlooking their reliability, particularly concerning issues like object hallucination. 
In our study, we introduce methods to assess the extent of object hallucination of publicly available LALMs. Our findings reveal that LALMs are comparable to specialized audio captioning models in their understanding of audio content, but struggle to answer discriminative questions, specifically those requiring the identification of the presence of particular object sounds within an audio clip. 
This limitation highlights a critical weakness in current LALMs: their inadequate understanding of discriminative queries. 
Moreover, we explore the potential of prompt engineering to enhance LALMs' performance on discriminative questions.

\end{abstract}

\section{Introduction}
Large audio-language models (LALMs) augment traditional large language models (LLMs) by incorporating audio perception capabilities. These models can accept both audio and text instructions as inputs, facilitating a broader spectrum of audio-related tasks. Recently, numerous works \cite{gong_ltuas, gong2023listen, chu2023qwen, tang2023salmonn, kong2024audio, shukor2023unified, shu2023llasm, zhang2023video, lyu2023macaw, chen2023lauragpt, liu2023music, wu2023decoder, deshmukh2024pengi, wang2023slm, pan2023cosmic, wu2023next, zhan2024anygpt} have proposed integrating audio perception modules and LLMs into a single multimodal model, which enables LLMs to process and understand audio inputs and handle various speech and audio tasks.

In the field of evaluation for LALMs, Dynamic-SUPERB \cite{huang2023dynamic} is introduced as the first benchmark focusing on speech processing tasks through designed questions and answer options.
Following this, AIR-Bench \cite{yang2024air} was developed, employing open-ended questions with GPT-4 \cite{openai2023gpt} as the evaluator to assess LALMs' task performance.
Both benchmarks primarily evaluate task performance but do not adequately assess the reliability of the content generated by LALMs, particularly concerning object hallucination. 
The issue of hallucination \cite{zhang2023siren, huang2023survey, rawte2023survey} has been raised with the rapid development of LLMs, and numerous works \cite{li2023evaluating, wang2023evaluation, zhou2023analyzing, wang2023llm, lovenia2023negative, dai2023plausible, zhai2023halle} in the field of computer vision have observed that large vision-language models (LVLMs) exhibit significant object hallucination in image captioning tasks, generating captions that include objects not present in the images.

Given the lack of discussion on object hallucination in LALMs within speech and audio domains, and the absence of benchmarks specifically measuring object hallucination, this paper introduces discriminative and generative tasks aimed at exploring the object hallucination phenomenon within LALMs.


Discriminative tasks aim to ascertain the presence of a specific object's sound within an audio clip by asking the model questions such as, ``Is there a sound of a dog in the audio?''.
The design of these questions involves sampling objects through both positive and negative methods.
Consequently, we can treat discriminative tasks as a binary classification task, calculating metrics such as accuracy, precision, recall, and the F1 score to evaluate performance. 
On the other hand, generative tasks are designed to guide models in performing audio captioning tasks through instructions, such as ``Describe the audio''.
After obtaining the model's predicted audio captions, nouns are extracted from these captions using NLP tools.
These nouns are considered as objects present in the sound, which are then compared with the ground truth labels to ascertain the extent of object hallucination exhibited by the model.

In the results of these two different types of tasks, we found that LALMs suffer from object hallucination and tend to give affirmative answers.
In addition, the performance of these LALMs is highly sensitive to prompt design.
Interestingly, LALMs rival specialized models in audio captioning tasks, demonstrating their ability to comprehend audio information.
However, their performance in discriminative tasks is less satisfactory. Although the models are adept at audio captioning, they struggle with answering discriminative questions and suffer from object hallucination.
We observe that LALMs struggle to understand discriminative questions, failing to extract the required information from the given audio.
Thus, we propose methods to improve their performance on discriminative tasks.
Our contributions are outlined as follows:
\begin{itemize}
    \item This is the first work to explore object hallucination in large audio-language models (LALMs).
    \item We observe that LALMs perform well on audio captioning tasks but struggle with answering discriminative questions, and we propose methods for improvement.
\end{itemize}

\begin{figure*}[t]
\centering
\includegraphics[width=1.00\textwidth]{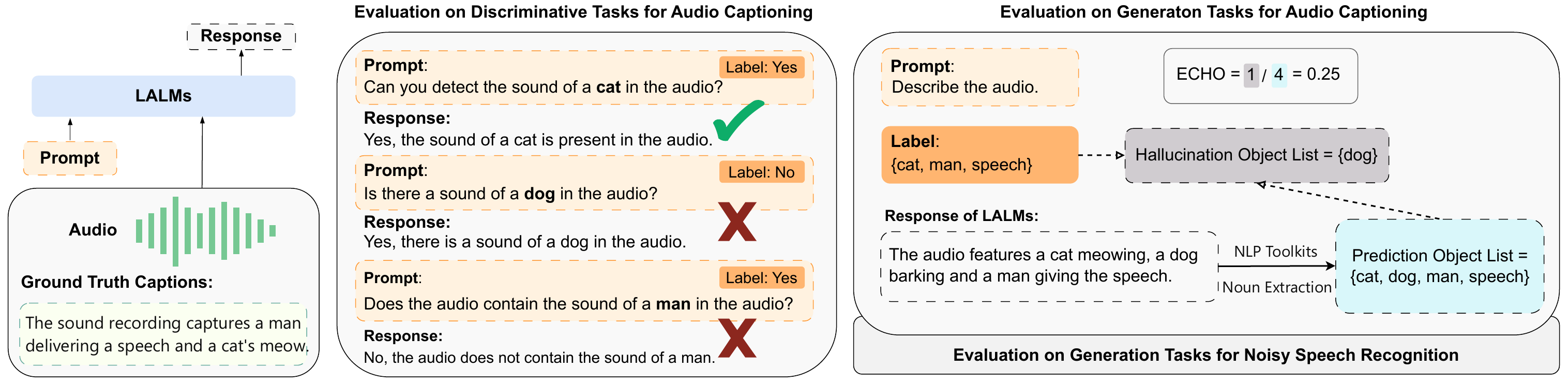}
\caption{Demonstration of our evaluation pipeline.}
\label{evaluation-pipeline}
\end{figure*}

\section{Evaluation Methods}

    

    In Sec \ref{sec:evaluation-methods-dataset}, we introduce the dataset. Sec \ref{sec:evaluation-methods-discriminative-tasks} and \ref{sec:evaluation-methods-generative-tasks} detail evaluation methods and metrics for discriminative and generative tasks, respectively, as further illustrated in Figure~\ref{evaluation-pipeline}.

\subsection{Dataset}
\label{sec:evaluation-methods-dataset}
    \begin{itemize}
        \item \textbf{Audio}:
        AudioCaps \cite{kim2019audiocaps} is designed for audio captioning tasks. It is 
        distinguished by its emphasis on human-annotated captions for a wide 
        variety of audio clips from the AudioSet \cite{7952261}, which features labels 
        for different sound events and categories identified within these clips. 
        Therefore, each audio clip is accompanied by text captions and multiple 
        labels.

        \item \textbf{Speech}:
        CHIME-6 \cite{watanabe2020chime} is tailored for advancing ASR systems in noisy conditions. 
        We use it for noisy speech recognition to validate the models' capabilities in ASR under conditions with environmental noise interference.
        To ensure that the transcriptions feature a sufficient number of nouns, we selected instances in the test set with more than three nouns in the transcription, totaling 489 instances.
        
    \end{itemize}

\subsection{Discriminative Tasks}
\label{sec:evaluation-methods-discriminative-tasks}

    \textbf{Methods}: Inspired by Polling-based Object Probing Evaluation 
    (POPE) \cite{li2023evaluating}, we adopt the similar approach for evaluating 
    LALMs.
    We formulate the evaluation of object hallucination as a binary 
    classification task that prompts LALMs to output ``Yes'' or ``No''. 
    We design five different prompts, ``Is there a sound of [object]?'', ``Does the audio contain the sound of [object]?'', `Have you noticed the sound of [object]?'', ``Can you hear the sound of [object]?'' and ``Can you detect the sound of [object]?''.
    In this way, by utilizing different strategies to sample objects that LALMs prone to hallucinate, we can establish a set of questions to poll LALMs. Since the expected answers to these discriminative questions are simply ``Yes'' or ``No'', we can easily 
    identify them without complicated parsing rules.
    
    Questions with answers are ``Yes'' can be directly built from ground truth 
    objects, while questions whose answers are ``No'' can be built by sampling from 
    negative objects. 
    Hence, we can devise various sampling strategies to validate whether LALMs are prone to hallucinate the specific objects.
    We consider all the ground truth labels for each audio as 
    positive samples. 
    Inspired by \cite{li2023evaluating}, we utilize the following negative sampling strategies:
    
    \begin{itemize}
        \item Random Sampling: We random sample {\it k} objects that are not present 
        in the current audio.
        \item Popular Sampling: We select the top-{\it k} most frequent objects across the entire audio captioning dataset that are not present in the current audio.
        \item Adversarial Sampling: We rank all objects by their frequency of co-occurrence with the actual objects in the audio. Subsequently, we select the top-{\it k} objects with the highest co-occurrence rates that are not included in the audio.
        
    \end{itemize}
    
    To ensure a balanced ratio between positive and negative samples during data construction, the parameter {\it k} is adjusted to match the number of ground truth labels associated with each audio clip. 
    Hence, we derive 15,110 positive instances directly from the ground truth.
    For each negative sampling strategy, we sample the equivalent number of negative instances, which is 15,110, to maintain this balance.
    \\
    
    \noindent\textbf{Metrics}:
    We adopt accuracy, 
    precision, recall and F1 score as the evaluation metrics.
    Because the evaluation is aimed at hallucination, both precision and recall are 
    calculated in relation to hallucination questions, where the ground truth answer is ``No''. 
    In addition, we report the ratio that LALMs answer ``Yes'' as a 
    reference to analyze the behavior of models.


\begin{table*}[ht]
    \scriptsize
    \centering
    \setlength{\tabcolsep}{9.5pt}
    \caption{Results of discriminative tasks for audio captioning: Acc (Accuracy), P (Precision), R (Recall), F1 (F1 scores), and Std (standard deviation of F1 scores across five prompts).
    Greedy and Sample refer to the decoding strategies.
    (Unit: \%)}
    \begin{tabular}{@{}clcccccccccccc@{}}
    \toprule
    \multicolumn{1}{c}{} & \textbf{} & \multicolumn{6}{c}{\footnotesize{Sample}} & \multicolumn{6}{c}{\footnotesize{Greedy}} \\
    \cmidrule(lr){3-8} \cmidrule(lr){9-14}
    \multicolumn{1}{c}{\textbf{POPE}}
    & \textbf{Model} 
    & \textbf{Acc} & \textbf{P} & \textbf{R} & \textbf{F1} & \textbf{Yes} & \textbf{Std} 
    & \textbf{Acc} & \textbf{P} & \textbf{R} & \textbf{F1} & \textbf{Yes} & \textbf{Std} 
    \\
    \midrule
    \multirow{5}{*}{Random} 
    & Qwen-Audio-Chat-7B 
    & {65.3} & 79.2 & 32.8 & 46.1 & 79.3 & 12.9
    & 65.0 & \textbf{97.0} & 30.5 & 46.4 & 84.3 & 23.0
    \\
    & LTU-AS-7B 
    & 50.1 & 49.2 & {46.5} & {47.8} & 52.8 & 5.2
    & 51.7 & 52.0 & {43.9} & {47.7} & 57.8 & 16.8
    \\
    & SALMONN-7B 
    & 56.3 & 90.0 & 14.1 & 24.4 & 92.2 & 29.5
    & 56.0 & 89.9 & 13.4 & 23.4 & 92.5 & 29.8
    \\
    & SALMONN-13B 
    & 63.7 & \textbf{95.7} & 28.6 & 44.1 & 85.1 & 25.7
    & {67.6} & 96.6 & 36.6 & 53.1 & 81.1 & 15.3
    \\
    & Specialized-LLaMA 
    & 65.5 & 60.9 & 82.7 & 70.1 & 32.0 & 3.1
    & 67.5 & 61.9 & 88.6 & 72.9 & 28.4 & 3.6
    \\
    & Specialized-ChatGPT
    & \textbf{77.1} & 69.3 & \textbf{96.6} & \textbf{80.7} & 30.3 & 0.7
    & \textbf{79.8} & 71.9 & \textbf{97.6} & \textbf{82.8} & 32.1 & 0.7
    \\
    \hline
    \addlinespace
    \multirow{5}{*}{Popular} 
    & Qwen-Audio-Chat-7B
    & 59.3 & 70.4 & 20.8 & 32.2 & 85.2 & 13.4
    & 58.1 & \textbf{96.8} & 16.8 & 28.4 & 90.7 & 25.3
    \\
    & LTU-AS-7B 
    & 47.1 & 45.6 & {40.5} & 42.9 & 55.6 & 5.2
    & 47.3 & 46.4 & {35.0} & 39.9 & 62.2 & 14.4
    \\
    & SALMONN-7B 
    & 56.8 & 90.4 & 15.2 & 26.0 & 91.6 & 27.3
    & 65.0 & 95.4 & 31.6 & 47.4 & 83.4 & 27.6
    \\
    & SALMONN-13B 
    & {65.2} & \textbf{96.0} & 31.7 & {47.6} & 83.5 & 23.4
    & {65.2} & 96.1 & 31.6 & {47.6} & 83.6 & 23.6
    \\
    & Specialized-LLaMA 
    & 58.1 & 55.4 & 68.0 & 61.1 & 38.7
    & 5.0
    & 60.2 & 57.0 & 74.0 & 64.4 & 35.0
    & 2.0
    \\
    & Specialized-ChatGPT
    & \textbf{66.2} & 64.7 & \textbf{70.6} & \textbf{67.5} & 45.4 & 0.8
    & \textbf{66.5} & 65.1 & \textbf{70.9} & \textbf{67.8} & 45.5 & 0.6
    \\
    \hline
    \addlinespace
    \multirow{5}{*}{Adversarial} 
    & Qwen-Chat-7B 
    & 57.6 & 70.7 & 16.6 & 26.8 & 88.4 & 12.4
    & 56.0 & 89.5 & 11.8 & 20.8 & 93.5 & 19.7
    \\
    & LTU-AS-7B 
    & 49.0 & 47.4 & 44.2 & {45.8} & 53.8 & 4.7
    & 50.1 & 49.6 & {40.6} & {44.6} & 59.5 & 15.8
    \\
    & SALMONN-7B 
    & 55.2 & 87.5 & 11.2 & 19.9 & 93.6 & 22.4
    & 60.3 & 93.3 & 21.3 & 34.7 & 88.7 & 22.7
    \\
    & SALMONN-13B 
    & {60.4} & \textbf{94.3} & {21.3} & 34.7 & 88.8 & 18.3
    & {60.4} & \textbf{94.3} & 21.3 & 34.8 & 88.8 & 18.5
    \\
    & Specialized-LLaMA 
    & 59.8 & 56.6 & 71.6 & 63.2 & 37.3
    & 5.2
    & 62.7 & 58.6 & 79.3 & 67.4 & 32.9
    & 1.6
    \\
    & Specialized-ChatGPT
    & \textbf{70.4} & 66.6 & \textbf{79.2} & \textbf{72.3} & 41.1 & 0.4
    & \textbf{70.7} & 67.1 & \textbf{79.6} & \textbf{72.8} & 41.2 & 0.4
    \\
    \bottomrule
    \end{tabular}
    \label{tab:discriminative_tasks}
\end{table*}

\begin{table*}[ht]
    \scriptsize
    \centering
    \caption{Results of generative tasks. 
    E stands for ECHO, with subscripts I and S indicating instance and sentence level, respectively. 
    The subscript g refers to the use of GPT-4 as the evaluator. Std denotes the standard deviation of ECHO across five prompts. 
    SPICE is referenced from \cite{anderson2016spice}.
    Due to its reliance on fixed template inputs, Qwen-Audio cannot accommodate varying prompts. 
    Specialized in audio captioning refers to specialized audio caption model \cite{kadlčík2023whisper}, while in ASR, it refers to Whisper \cite{radford2023robust} large v3.
    (Unit: \%)
    } 
    \begin{tabular}{@{}clcccccccccccccc@{}}
    \toprule
    \multicolumn{1}{c}{} & \textbf{} & \multicolumn{9}{c}{\footnotesize{Audio Captioning}} & \multicolumn{4}{c}{\footnotesize{Noisy Speech Recognition}} \\
    \cmidrule(lr){3-11} \cmidrule(lr){12-15}
    \multicolumn{1}{c}{\textbf{Strategy}} 
    & \textbf{Model}
    & \textbf{E$_{\text{I}}$} $\downarrow$ 
    & \textbf{E$_{\text{S}}$} $\downarrow$
    & \textbf{Cov} $\uparrow$
    & \textbf{Std$_{\text{I}}$}
    & \textbf{Std$_{\text{S}}$}
    & \textbf{E$_{\text{I,g}}$} $\downarrow$ 
    & \textbf{E$_{\text{S,g}}$} $\downarrow$ 
    & \textbf{Cov$_{\text{g}}$} $\uparrow$ 
    & \textbf{SPICE } $\uparrow$ 
    & \textbf{E$_{\text{I}}$} $\downarrow$ 
    & \textbf{E$_{\text{S}}$} $\downarrow$ 
    & \textbf{Cov} $\uparrow$ 
    & \textbf{WER} $\downarrow$ 
    \\
    \midrule
    \multirow{5}{*}{Sample} 
    & Qwen-Audio-Chat-7B
    & 39.5 & 66.3 & 13.1 
    & 0.9 & 1.4
    & 25.7 & 44.0 & 14.6
    & 22.2 
    & 25.3 & 76.6 & {56.6} & 44.0
    \\
    & Qwen-Audio-7B 
    & 38.0	& 58.6 & 11.5 
    & - & -
    & 25.8 & 45.0 & 3.4
    & 11.1 
    & {17.3} & 77.6 & 36.3 & \textbf{30.3}
    \\
    & LTU-AS-7B 
    & 85.2 & 90.0 & 2.1 
    & 8.7 & 9.5
    & 84.1 & 93.9 & 0.8
    & 5.3 
    & 47.6 & {53.2} & 56.1 & 97.0
    \\
    & SALMONN-7B 
    & \textbf{33.5} & {57.1} & \textbf{15.2} 
    & 15.2 & 19.2
    & \textbf{20.7} & \textbf{36.7} & 18.8
    & \textbf{23.4} 
    & 30.9 & 62.6 & 44.6 & 73.3
    \\
    & SALMONN-13B 
    & 43.6 & 69.1 & 12.7 
    & 15.6 & 21.4
    & 22.7 & 40.2 & \textbf{19.2}
    & 21.5 
    & 32.9 & 61.0 & 46.6 & 67.0
    \\
    & Specialized
    & \textbf{33.5} & \textbf{56.5} & 13.2
    & - & -
    & 26.2 & 44.6 & 15.4
    & 13.2
    & \textbf{16.8} & \textbf{33.2} & \textbf{74.9} & 37.0
    \\
    \hline
    \addlinespace
    \multirow{5}{*}{Greedy} 
    & Qwen-Audio-Chat-7B 
    & 27.8 & 50.3 & \textbf{15.1}
    & 2.1 & 5.7
    & \textbf{20.4} & 35.0 & 16.7
    & \textbf{22.2}
    & 18.5 & 42.4 & 80.0 & {31.0}
    \\
    & Qwen-Audio-7B 
    & \textbf{25.4} & \textbf{38.5} & 12.5 
    & - & -
    & 20.6 & \textbf{33.9} & 17.1 
    & 13.7 
    & {14.2} & 31.1 & {80.2} & 41.0
    \\
    & LTU-AS-7B 
    & 91.9 & 98.9 & 2.4
    & 4.1 & 3.1
    & 81.5 & 91.0 & 1.7
    & 6.0 
    & 14.8 & {27.3} & 79.3 & 32.0
    \\
    & SALMONN-7B 
    & 33.1 & 54.9 & 13.9 
    & 17.0 & 24.7
    & 21.9 & 37.6 & 18.8 
    & 21.87 
    & 61.0 & 68.6 & 34.5 & 73.0
    \\
    & SALMONN-13B 
    & 43.8 & 69.3 & 12.4
    & 15.7 & 23.0
    & 22.4 & 40.0 & \textbf{19.4}
    & 21.1 
    & 20.9 & 34.4 & 68.7 & 41.0
    \\
    & Specialized
    & 27.4 & 47.7 & 15.0
    & - & -
    & 24.5 & 40.5 & 18.2
    & 16.5 
    & \textbf{10.4} & \textbf{24.3} & \textbf{81.1} & \textbf{23.0}
    \\
    \bottomrule
    \end{tabular}
    \label{tab:generation_tasks}
\end{table*}

\begin{table*}[h]
    \small 
    \centering
    \caption{The \textbf{F1 scores} for discriminative tasks, under various prefix prompts and a random sampling strategy, are reported as differences relative to the baseline. This baseline is determined by the F1 scores from tests conducted without the addition of prefix prompts.
    P1: ``Listen.'', 
    P2: ``Listen closely to the audio before answering the following question.'', 
    P3: ``Carefully consider the question before providing your answer.'', 
    P4: ``Listen closely to the audio and carefully consider the question before providing your answer.'',
    P5: ``Focus and answer the following question.'', 
    P6: ``Focus on the given audio and answer the following question.'', 
    P7: ``Focus on the question and provide the answer.'', 
    P8: ``Focus on the given audio and the question and provide the answer.''
    (Unit: \%)
    }
    \begin{tabular}{@{}llcccccccc@{}}
    \toprule
    & \textbf{Model} 
    & \textbf{P1} 
    & \textbf{P2} 
    & \textbf{P3} 
    & \textbf{P4} 
    & \textbf{P5} 
    & \textbf{P6} 
    & \textbf{P7} 
    & \textbf{P8} 
    \\
    \midrule
    \multirow{5}{*}{}
    & Qwen-Audio-Chat-7B 
    & -11.2
    & +5.7
    & +12.8
    & +9.0

    & +6.1
    & +12.1
    & 0.0
    & +6.8
    
    \\
    & LTU-AS-7B 
    & -9.4
    & -8.2
    & -5.8
    & -16.5

    & -6.8
    & -12.3
    & -7.1
    & -8.6
    \\
    & SALMONN-7B 
    & +5.3
    & +14.7
    & +28.4
    & +30.4

    & +14.7
    & +32.0
    & +22.5
    & +32.2

    \\
    & SALMONN-13B
    & +3.7
    & +8.6
    & +27.0
    & +25.1
    
    & +7.0
    & +17.4
    & +12.9
    & +5.7
    \\
    \bottomrule
    \end{tabular}
    \label{tab:prefix_prompt}
\end{table*}

\subsection{Generative Tasks}
\label{sec:evaluation-methods-generative-tasks}
    \textbf{Methods}:
    We devise five different prompts, ``Describe the audio'', ``What do you hear?'', ``What can be inferred from the audio?'', ``This is a sound of?'' and ``Generate audio caption:'',
    to prompt LALMs to generate 
    captions for given audio clips. 
    Next, we use NLP tools, SpaCy \cite{spacy2}, to extract 
    nouns from these captions, identifying these nouns as the objects that the model perceives to be producing sounds in the audio. This approach allows us to compile a list of objects identified by the LLM as present in the audio.
    On the other hand, AudioCaps \cite{kim2019audiocaps} dataset provides 
    corresponding ground truth captions and labels (lists of objects producing sounds 
    in the audio). By applying 
    the same method to extract nouns from these ground truth captions and then 
    combining them with the ground truth labels, we can obtain a 
    comprehensive list of ground truth objects. 
    By employing the same process, we can adapt it to tasks involving noisy automatic speech recognition, substituting the prompt with ``What spoken text can be heard?''. 
    The ground truth labels are then the objects within the transcriptions.
    \\
    
    \noindent\textbf{Metrics}:
    We propose two metrics for evaluating generative audio captioning tasks.
    First, similar to the concept of Caption Hallucination Assessment with Image 
    Relevance (CHAIR) \cite{rohrbach2018object}, we propose the metric named ECHO, which is Evaluation of 
    Caption Hallucination in audiO, to evaluate object hallucination in audio 
    captioning tasks. 
    Given the ground truth objects in the audio, ECHO calculates 
    the proportion of objects appearing in the caption but not being present in the 
    audio. 
    We adopt two variants, ECHO$_{\text{I}}$ and ECHO$_{\text{S}}$, to evaluate hallucination at the object instance and sentence levels, respectively.
    They can be expressed as:

    \begin{equation}
        \small
        \text{ECHO}_I = \frac{\left|\{\text{hallucinated objects}\}\right|}{\left|\{\text{mentioned objects in the caption}\}\right|},
    \end{equation}

    \begin{equation}
        \text{ECHO}_S = \frac{\left|\{\text{captions with hallucinated objects}\}\right|}{\left|\{\text{all captions}\}\right|}.
    \end{equation}

    Second, Cover (Cov) measures the extent to which response cover the audio captions. It can be expressed as:
    \begin{equation}
        \text{Cover} = \frac{\left|\{\text{objects in the caption that actually exist in audio}\}\right|}{\left|\{\text{ground truth objects}\}\right|}.
    \end{equation}

    In order to mitigate potential errors introduced by automated measurement methods, we introduce GPT-4 as a reference baseline.
    We feed captions generated by LALMs, along with ground truth captions and labels, into GPT-4, requesting it to analyze which objects are considered hallucinations and which are not.
    Consequently, we can also 
    use the decomposed results of GPT-4 and calculate the ECHO and Cover 
    scores. We name it as ECHO$_{\text{I, g}}$, ECHO$_{\text{S, g}}$ and Cover$_{\text{g}}$. 
    For noisy automatic speech recognition task, we also report the word error 
    rate (WER) as a reference and select Whisper \cite{radford2023robust} large v3 as a baseline model for comparison.


    

\subsection{Evaluation Settings}

    We select five publicly available LALMs: Qwen-Audio \cite{chu2023qwen}, 
    Qwen-Audio-Chat \cite{chu2023qwen}, LTU-AS-7B \cite{gong_ltuas}, SALMONNN-7B 
    \cite{tang2023salmonn} and SALMONNN-13B \cite{tang2023salmonn}.
    We explore greedy and sample decoding strategies, setting the temperature to 1.0, top p to 0.9, and top k to 50 for sample-based decoding. Sample decoding is performed three times to calculate the average results.
    In addition, as an intuitive baseline, we cascade an existing Whisper-based audio captioning model \cite{kadlčík2023whisper} with ChatGPT (gpt-3.5-turbo-0125) or LLaMA-7b-chat \cite{touvron2023llama} to serve as a reference, which is named as Specialized-ChatGPT and Specialized-LLaMA, respectively.
    For cascade pipeline, we combine captions obtained from the audio captioning model and transcriptions acquired through Whisper with the original discriminative questions as input to LLMs to obtain responses. 
    Our codes are available at:
    \textcolor{deepblue}{{\scriptsize\url{github.com/kuan2jiu99/audio-hallucination}}}.

\section{Evaluation Results}

\subsection{Results on Discriminative Tasks}

    Illustrated in Table~\ref{tab:discriminative_tasks}, we observe that the recall scores of all LALMs are significantly lower than their precision scores. This suggests that LALMs tend to provide affirmative answers when addressing the issue of hallucination, indicating that LALMs are easily misled by non-existent objects. Given that we maintained a 1:1 ratio between ground truth and non-existent objects, the results from the yes rate reveal that most LALMs are predisposed to giving a ``Yes'' response. 

    Second, due to the differences in sampling strategies, the difficulty of answering questions varies for the models, resulting in most LALMs experiencing a decrease in F1 scores according to the Random, Popular, and Adversarial settings. From this, it is evident that LALMs are more prone to hallucinating about objects that appear frequently or concurrently.
    Additionally, in Table~\ref{tab:discriminative_tasks}, we report the standard deviation of the F1 scores for different prompts, highlighting the models' sensitivity to various prompt designs. We also discover that the prompt ``Have you noticed the sound of [object]?'' consistently yields notably high F1 score performance across all models.
    Third, F1 score of the cascade pipeline significantly surpasses that of all LALMs, indicating a substantial gap between current LALMs and cascade pipelines that needs to be bridged. 

\subsection{Results on Generative Tasks}

    Illustrated in Table~\ref{tab:generation_tasks}, the performance of LALMs on the ECHO and Cover metrics is comparable to that of Whisper-based caption model \cite{kadlčík2023whisper}.
    The results obtained using GPT-4 as evaluator exhibit consistent trends.
    This indicates that LALMs are capable of generating high-quality audio captions, demonstrating sufficient ability to understand audio information. 
    Furthermore, the extent of object hallucination is similar among them except for LTU-AS, suggesting that LALMs can match the performance of Whisper-based audio captioning models in both understanding audio content and the level of object hallucination.
    Compared to the significant performance gap on discriminative tasks between LALMs and cascade pipelines, 
    it is evident that even though LALMs exhibit performance that rivals specialized caption models in audio captioning tasks, they falter significantly when faced with discriminative questions.
    This discrepancy highlights a specific weakness of LALMs in handling tasks that require precise discrimination based on audio content, despite their competency in generating descriptive captions from audio inputs.
    Since LALMs are capable of understanding information within audio, it suggests that the issue may not lie with their ability to process audio content. 
    Instead, it is likely that LALMs struggle to fully comprehend the nature of discriminative questions, making it challenging for them to extract the required information from the given audio.
    Besides, compared to discriminative tasks, the greedy decoding strategy effectively reduces hallucination phenomena in generative tasks.

\subsection{Results on Prompt Engineering}

    Building on the observation that LALMs are indeed capable of understanding audio information, suggesting their struggle is not with processing audio content but with fully comprehending the nature of discriminative questions, we add appropriate prefix prompts before discriminative questions.
    These prefix prompts are denoted as P1 to P8, as shown in Table \ref{tab:prefix_prompt}, where F1 scores are reported as differences relative to the baseline F1 score. The baseline score is the result without the addition of prefix prompts.
    By doing so, we anticipate that LALMs thereby focus on seeking the specific information required from the audio to address the question.
    For P1 to P4, by emphasizing the importance of listening to the given audio, contemplating the question, or both, we experimented with the impact on LALMs' performance. 
    In Table \ref{tab:prefix_prompt}, we discover that instructing LALMs to carefully consider the question and to both attentively listen to the audio and carefully consider the question yielded the most significant improvement in performance except for LTU-AS.
    Merely emphasizing listening to the given audio do not achieve as substantial an improvement as instructing the model to carefully consider the question.
    On the other hand, for P5 to P8, by emphasizing focus on the given audio, the question, or both, we examine their effects on LALMs' performance.
    Our findings indicate that instructing LALMs to concentrate on specific information or both can enhance performance except for LTU-AS. 
    This suggests the importance of clearly directing LALMs towards the focus of their attention.
    Additionally, we experimented with substituting "Focus" with synonyms like "Pay attention" and "Concentrate", yielding similar results.
    Conclusively, designing an appropriate prompt significantly impacts the performance on discriminative questions.

\section{Conclusion and Future Work}

    Despite recent advancements, LALMs present reliability concerns, particularly with object hallucination.
    We deploy task-oriented methodologies to gauge this issue, revealing that despite LALMs' capability to perform audio captioning comparably to specialized models, they struggle significantly with discriminative tasks and exhibit severe object hallucination. 
    Compared to cascade pipelines, LALMs still have considerable ground to cover in addressing object hallucination challenges.
    We propose improvement methods, whereby instructing LALMs to focus on specific information, listen to the audio, or carefully consider the question before responding can enhance model performance.
    Future work will refine prompts for LALMs to improve audio information extraction and response accuracy, and develop strategies to lessen hallucination in both pre-training and inference stages.

\section{Limitations}
    
    In generative tasks, combining with automatic segmentation may lead to mismatches between extracted objects and human annotations, causing discrepancies. 
    Challenges also arise in noun extraction when LALMs inaccurately generate captions. 
    To mitigate these issues, we also utilize GPT-4 as our evaluator.
    In discriminative tasks, LALMs occasionally ignore instructions, complicating automated evaluation by not providing binary answers.

\section{Acknowledgement}
We would like to thank En-Pei Hu and Ke-Han Lu for providing valuable feedback on the draft of this paper. 
Additionally, we thank the National Center for High-performance Computing (NCHC) of National Applied Research 
Laboratories (NARLabs) in Taiwan for providing 
computational and storage resources.

\bibliographystyle{IEEEtran}
\bibliography{mybib}

\end{document}